# Improved optomechanical disk resonator sitting on a pedestal mechanical shield

Dac Trung Nguyen[1], William Hease[1], Christopher Baker[1], Eduardo Gil-Santos[1],
Pascale Senellart[2], Aristide Lemaître[2], Sara Ducci[:1], Giuseppe Leo[1] and Ivan Favero[1]

[1] *Matériaux et Phénomènes Quantiques, Université Paris Diderot - CNRS, Sorbonne Paris Cité, 10 rue Alice Domon et Léonie Duquet, 75013 Paris, France.*

[2] *Laboratoire de Photonique et de Nanostructures, CNRS, route de Nozay, 91460 Marcoussis, France.*

**Abstract** We experimentally demonstrate the controlled enhancement of the mechanical quality factor $Q$ of GaAs disk optomechanical resonators. Disks vibrating at 1.3 GHz with a mechanical shield integrated in their pedestal show a $Q$ improvement by a factor 10 to 16. The structure is modeled numerically and different modes of vibration are observed, which shed light on the $Q$ enhancement mechanism. An optimized double-disk geometry is presented that promises $Q$ above the million for a large parameter range.

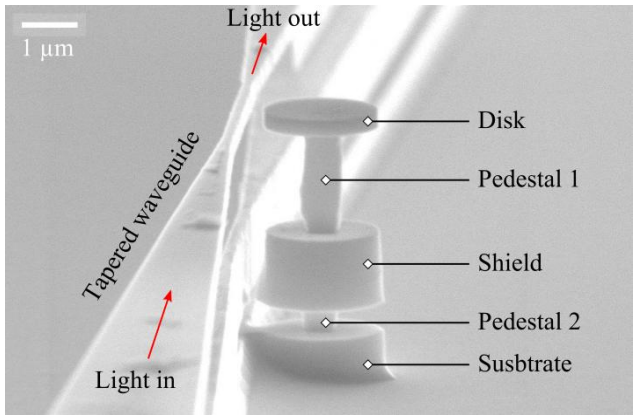

*Fig. 1: Scanning Electron Microscope (SEM) image of a fabricated shielded optomechanical disk resonator with its suspended optical coupling waveguide.*

## Introduction

Optomechanical techniques offer the possibility to manipulate the motion of miniature mechanical systems by optical means [1-3]. Amongst the large variety of optomechanical settings that are by now investigated, Gallium Arsenide (GaAs) disk resonators combine excellent features like a strong coupling between optical Whispering Gallery Modes (WGM) and radial breathing motion (RBM), a small footprint, and perspectives of optoelectronic integration with optically active elements [4]. In dwarf disk of radius around a micron, the single photon optomechanical coupling $g_0$ lies in the few MHz range [5,6], which allows a high degree of optical control over GHz-frequency RBMs.

On the mechanical side, a high quality factor $Q$ is key to the high precision mechanical detection of signals, to the classical or quantum coherence of motion, and to the performance of some MEMS applications [7]. At cryogenic temperature and under vacuum, the $Q$-frequency product of state of the art GaAs disks attains $10^{13}$ and appears to be presently limited by clamping losses [8]. Hence an optimization of the geometrical design is expected to lead to substantial $Q$ improvement [8-12]. In a previous work, we introduced the principle of a mechanical shield integrated into the disk's pedestal to enhance the mechanical $Q$ [8]. The idea is to reduce clamping losses by coupling the vibration modes of the disk and the shield destructively, in order to reduce vibration amplitude at the pedestal/substrate anchoring point.

Here we report on the fabrication and characterization of such shielded pedestal geometry. With a not yet fully optimized hetero-structure and disks of pedestal radius larger than 250 nm, we already demonstrate a boost of $Q$ by more than a factor 10 at GHz frequencies, reaching $Q$ of about two thousands and proving the validity of the approach. While we have obtained similar values of $Q$ without shield for pedestal's radius smaller than 120 nm [8], the shielded pedestal concept will lead to further improvements. As an example, we introduce in the last section a compact double-disk structure with expected $Q$ above the million.

## Sample fabrication

| Layer | Thickness (nm) |
|---|---:|
| GaAs | 320 |
| $Al_{0.8}Ga_{0.2}As$ | 1800 |
| GaAs | 1250 |
| $Al_{0.8}Ga_{0.2}As$ | 450 |
| GaAs | ~500 μm substrate |

*Tab. 1: Epitaxial structure of the wafer used for the shielded GaAs disk resonators. The physical parameters of GaAs and $Al_{0.8}Ga_{0.2}As$ employed in this work are taken from Refs. [13-16].*

The shielded GaAs disk resonators are fabricated from an undoped GaAs/AlGaAs multilayer grown on a GaAs semi-insulating substrate (Tab. 1). Disks of radius 1.05 µm are positioned in the vicinity of GaAs suspended tapered optical waveguides, allowing evanescent guide-to-disk optical coupling (Fig. 1). The disks and waveguides are defined in a resist mask (ma-N 2405 from *micro resist technology*) by electron beam lithography, followed by non-selective inductively coupled plasma reactive ion etching (ICP-RIE) using a mixture of $SiCl_4$ and Argon plasmas. Pedestals are formed by selective under-etching of the AlGaAs sacrificial layers, using hydrofluoric acid (HF). Remaining resist is removed by a thorough oxygen plasma cleaning.

The ICP-RIE step is critical in the fabrication process of these high aspect-ratio resonators. Indeed, the narrowest fabricated features reach 200 nm (width of the waveguide tapered part and disk-guide lateral gap distance) when at the same time, the etch depth must be at least 3.5 µm in order to release the shield from the nearby parts (see Fig. 1). Because of the RIE lag [17,18] (inhomogeneous plasma etch rate due to the inhomogeneous size of the smallest patterns open in the mask), we must etch the largest patterns about 5 µm in depth, which means an etch selectivity of 10:1 for the 500 nm thick resist mask. We developed a multi-step ICP procedure to reduce the impact of RIE lag and obtain vertical sidewalls. The sample is dry-etched in successive steps of 5 minutes. After each step, plasma sources are shut down and the chamber pumped out to evacuate residual inactive substances in the narrowest zones on the sample. 35 to 40 steps are necessary to obtain this way the desired etch depth of about 5 µm.

Fig. 1 shows an SEM image of a fabricated shielded disk resonator, with an etch depth of 5 µm on the right side of the disk and 3.5 µm in the disk-waveguide gap. The proper ICP-RIE and HF etching of these structures is extremely challenging, such that the final geometry slightly departs from the nominal design. For instance, the pedestal is not perfectly centered under the disk, which will play a role in the resonator's mechanical properties as we will see in the following section. The shield pedestal is barely separated from a nearby wall-like structure sitting vertically below the waveguide defined in the top GaAs layer. This wall, which reproduces the tapered profile of the top waveguide, does not play any role in the optomechanical properties of the resonators discussed in this work.

**Optomechanical experiments**

In this study, we measure the mechanical Brownian motion of GaAs disk resonators by optomechanical means [4,5]. CW laser light at wavelength λ~1.3 µm is coupled into the waveguide, propagates, couples evanescently to the disk's optical modes and interacts with the disk's mechanical motion [19]. The optical output signal passes a semiconductor optical amplifier (SOA) before impinging on a fast photodetector whose electrical output RF spectrum is acquired with a spectrum analyzer.

The laser is tuned to the blue flank of a WGM resonance of the disk in order to reveal its mechanical resonances in the RF spectrum. We employ the largest detuning and lowest optical power possible, such that optomechanical dynamical back-action is completely negligible and does not lead to over-estimation of the mechanical $Q$. In the present experiment we evaluate that $Q$ is measured with ±15% error. Fig. 2a shows an example of a mechanical spectrum obtained by such optomechanical measurement, in a cryostat operated at pressure ≤$10^{-4}$ mbar. The resonant peak frequency at 1.357 GHz corresponds to the first order radial breathing mode (RBM) of the disk. $Q$ is extracted by fitting this spectrum with a Lorentzian profile atop a constant offset.

To reach higher values of $Q$ we progressively reduce the pedestal radius $r$ by increasing the duration of the HF underetch. The fragility and the vertical geometrical irregularity of the structure sets a limit $r \geq 250$ nm for current shielded disks. In Fig. 2b, the measured $Q$ is plotted as a function of the radius, both for shielded resonators (blue diamond) and for simple shield-less resonators (red circles, reproduced from Ref. [8]) for comparison. 2 to 5 disks were measured for each value of $r$. For shielded resonators, an improvement of $Q$ by a factor 10 to 16 is observed over the whole range of achieved pedestal radius. As fabricated shielded pedestals are not strictly cylindrical, an average value is quoted for their radius in Fig. 2b. A radius uncertainty is also inferred by measuring the departure from the perfect cylinder. The solid lines in the figure are obtained by Finite Element Method (FEM) numerical simulations of clamping losses [8], with the hashed area accounting for the radius uncertainty. Experimental data obtained with shielded structures fall within this area, suggesting that clamping losses are still the dominant dissipation mechanism even for these improved resonators. The observed 10-fold boost of $Q$ validates the pedestal shield working principles, promising enhanced performances in more optimized structures [8].

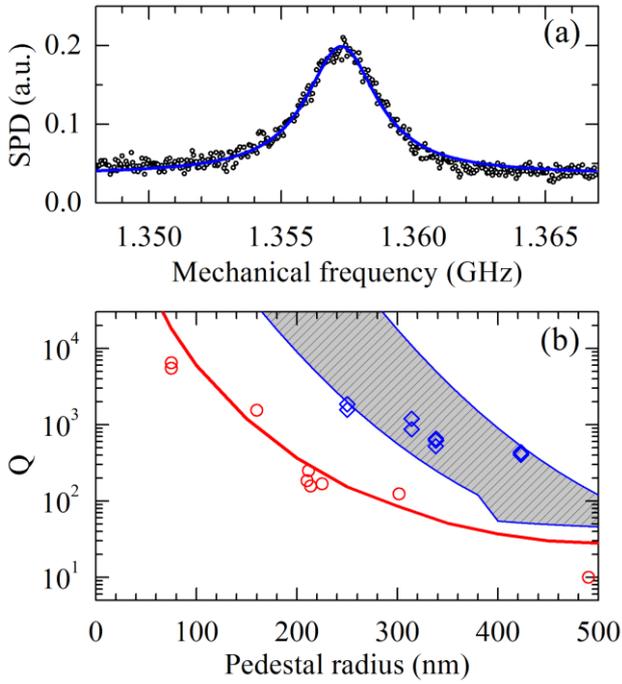

Fig. 2: Optomechanical measurement of the mechanical Q factor. (a) Example of a Brownian motion mechanical spectrum measured optically (black circles), fitted by a Lorentzian function (blue solid line), where SPD stands for spectral power density. (b) Mechanical Q versus pedestal radius r for simple shield-less disks (red, reproduced from Ref. [8]) and for shielded disks (blue). The hashed grey zone corresponds to simulated Q accounting for the 20% uncertainty on the pedestal radius.

Working at higher optical powers and varying the laser-cavity detuning gives insights into the mechanical behavior of shielded resonators. In Fig. 3a, an example of a mechanical spectrum obtained at high power and for the laser tuned to the middle of the blue flank is presented. This spectrum contains three visible peaks, in contrast to the Fig. 2a recorded at larger detuning. This difference is investigated by systematic variation of the detuning. In Fig. 3b, as the laser wavelength increases from 1330 to 1350 nm, going from large to small detuning on the blue flank, a mechanical peak appears first and is then joined by two additional peaks. As soon as the laser wavelength passes the thermally shifted optical resonance ($\lambda$=1350 nm, Fig. 3c), the system becomes unstable and mechanical peaks disappear from the spectrum.

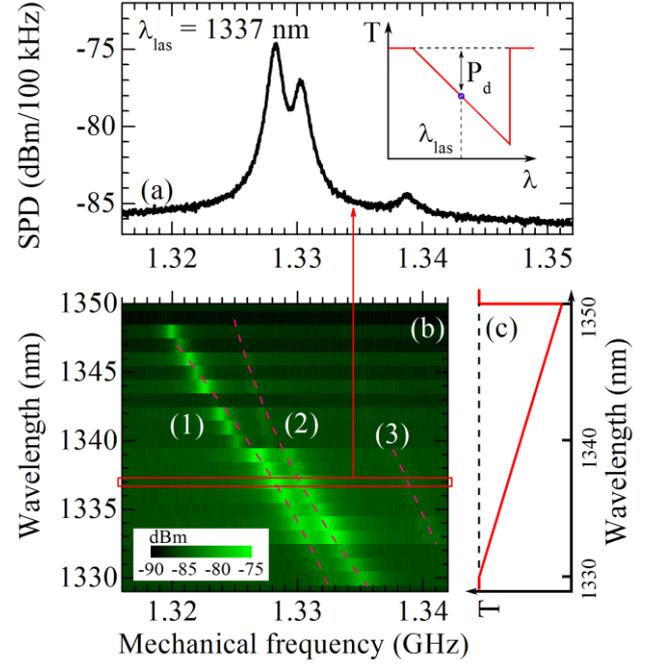

Fig. 3: Mechanical spectra at large power and different detunings. (a) Mechanical SPD for a laser wavelength $\lambda_{las}$ = 1337 nm. Inset: thermo-optical transmission spectrum of a disk, where $P_d$ denotes the power dropped in the disk. (b) SPD (log scale) for different laser detunings on the blue flank of the optical resonance. Dashed lines are guides to the eyes. (c) Illustration of the optical transmission in the thermo-optical regime, corresponding to the laser detunings employed in figure (b).

As apparent in Fig. 3b, the mechanical peaks shift to lower frequencies as we reduce the detuning, and they group into distinct dispersion lines corresponding to three mechanical modes. We observe a clear anticrossing between two of these. At largest detuning the mechanical energy is mostly concentrated in mode 2. As the detuning reduces, the weight of mode 2 diminishes while mode 1's weight increases gradually.

We will explain this behavior in terms of variation of the mechanical properties of GaAs disks with temperature. Starting from off-resonance blue detuning, an increase in laser wavelength corresponds to a decrease of the detuning and hence an increase of the optical power dropped into the disk $P_d$. The resonator's temperature rises as a result of residual absorption, and the mechanical frequencies decrease consequently. Indeed, elasticity theory for a disk composed of an isotropic elastic medium predicts a frequency for the first order RBM that scales like $f \propto 1/R\sqrt{E/\rho}$ with $R$ the disk radius, $E$ the Young modulus and $\rho$ the density [6,20]. These three parameters change with temperature and for small variation $\Delta T$ the frequency shift is given by $\Delta f/f = (-\alpha - \alpha_\rho/2 + \alpha_E/2)\Delta T$ where $\alpha$ [K$^{-1}$] is the linear thermal expansion coefficient, $\alpha_\rho$ [K$^{-1}$] the linear thermal density change and $\alpha_E$ [K$^{-1}$] the linear thermal change of Young modulus of GaAs. At room temperature, $\alpha = 5.7 \times 10^{-6}$ K$^{-1}$, $\alpha_\rho = -1.9 \times 10^{-5}$ K$^{-1}$ and $\alpha_E = -1.2 \times 10^{-4}$ K$^{-1}$ [16]. Hence the thermal frequency shift is dominated by the Young modulus variation.

As a 2-dimensional axisymmetric FEM approach only predicts the existence of the first order RBM in the frequency range of interest, we perform 3D numerical simulations in order to identify the three mechanical modes involved in experiments and their frequency shift versus temperature. For these simulations, the temperature field inside the entire shielded disk resonator is obtained by fixing a temperature increase $\Delta T$ at the disk's sidewall (modeling the heating by an optical WGM) and solving for stationary thermal state, taking into account the thermal change of the Young modulus. For each $\Delta T$, a number (5 to 10) of mechanical eigenmodes with frequencies around the first RBM are computed and analyzed. For the geometry employed in our experiments, two eigenmodes (1 and 2) are typically found to couple and display potential anticrossing in the frequency range of interest close to the RBM.

The anticrossing can be understood qualitatively as follows: the temperature rise in the resonator is a result of residual optical absorption, with the highest temperature reached at the disk's boundary where the optical WGM resides, and lower temperature in the shield. Because its deformation profile is concentrated within the top disk, the RBM frequency shifts thermally more than other modes, whose mechanical deformation lies mainly within the shield. Since modes 1 and 2 result from the coupling of the RBM with a secondary mode, they hence converge in frequency as the detuning is reduced and the temperature consequently increased (Fig. 3b), before the anti-crossing due to cross-coupling expels them one from another. In Fig. 3b, the shift of 15 MHz in the mechanical frequency when the laser wavelength passes from 1330 nm to 1350 nm corresponds to a temperature increase of about 200 K, obtained for the largest optical power employed in our experiments.

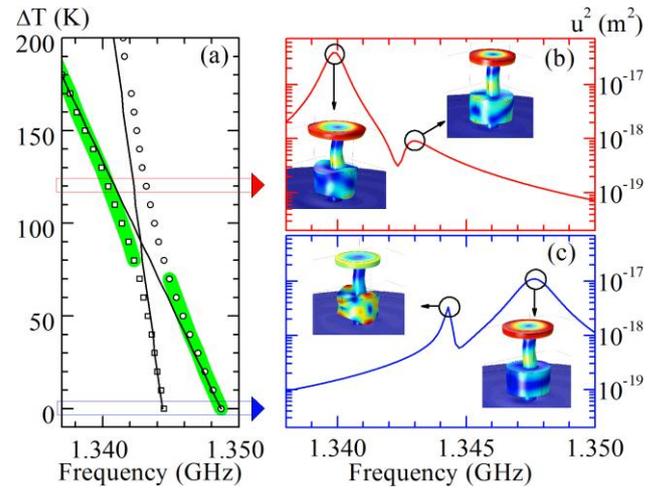

*Fig. 4: 3D simulation of a shielded disk resonator's modes as a function of temperature. (a) Mechanical frequency shifts for increasing temperature, for a resonator with perfectly centered pedestals (black solid lines) and for pedestals radially shifted by 80 nm (open black squares and circles). The green shadow highlights the mechanical mode with largest radial displacement (RBM-like). $\Delta T$ denotes the temperature rise at the disk's sidewall. (b) and (c) Squared radial displacement $u^2$ response spectrum at $\Delta T = 0$ and $\Delta T = 120$ K respectively.*

The geometric imperfection which is present in fabricated resonators accounts for some distinctive features of experimental results. Fig. 4a shows the obtained temperature shifts for mechanical frequencies of modes 1 and 2 in two cases: first for a resonator with perfectly centered pedestals (solid black lines); and second for a resonator with pedestals radially shifted by 80 nm, which is the value typically estimated on our finished shielded disks (black open symbols). From this figure, it is clear that mechanical modes do not anticross in the case of centered pedestals while they do for radially shifted pedestals. Hence, the anticrossing observed in Fig.3b seems to indicate coherent coupling of two mechanical modes of off-centered shielded structures, enabled by optically triggered thermal tuning. The resulting hybrid modes can, in case of optimized configurations, dramatically minimize displacement at the clamping point and enhance $Q$ [8].

For more details on mode shapes, Figs. 4b and 4c show 3D simulations performed on a disk with radially shifted pedestals, at the two temperatures highlighted in Fig. 4a. The solid lines are obtained through the simulated displacement response of the disk sidewall under a sinusoidal driving force, whose frequency is swept. In such simulations, the existence of spectrally close resonances can give rise to interferences and Fano profiles. The results shown in Figs. 4b and 4c illustrate the transfer of the RBM character (highlighted by green shadows in Fig. 4a) from one mechanical mode to the other as the temperature in the disk increases. This is equivalent to the transfer of signal weight observed in Fig. 3b as the detuning is reduced.

**Compact double-disk shielded structure**

The above shielded disk resonators are challenging to fabricate with no geometrical imperfection. First, deep ICP-RIE results in somewhat irregular pedestal and shield shapes, due to the RIE lag and the proximity between disk and waveguide. Second, the high aspect ratio involved makes the resonator fragile, presently preventing us from reaching smaller pedestal radius $r$ < 250 nm. To circumvent these difficulties, one idea is to reduce the total height of structure, in order to both mitigate the needed etch depth and obtain a more rigid structure. To that purpose, we develop a novel design where the vertical gap between the disk and the shield is reduced from 2 µm (in the above shielded disks) to 500 nm, while keeping optical losses low enough to preserve high optical $Q$ and high waveguide transmission. Besides, we also reduce the thickness of the shield from 1.35 µm to 300 nm. The resulting geometry is one with two disks of comparable but still distinct thickness, mechanically coupled through a common pedestal.

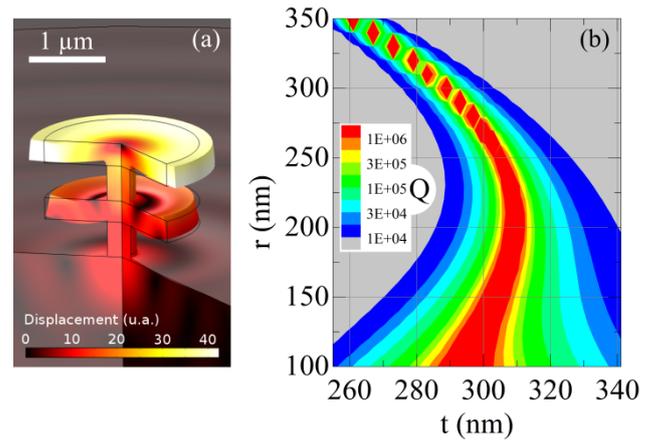

*Fig.5: Design of a GaAs double-disk resonator. (a) In the double-disk resonator shown here, the RBM of the top disk is in phase opposition with that of the bottom disk, leading for specific geometries to a displacement cancellation at the lower anchor. (b) Clamping-loss limited Q of a GaAs doubled-disk resonator top RBM as a function of the bottom disk's thickness* t *and pedestal radius* r*, numerically simulated using a 2D axisymmetric model.*

The design optimization of this double-disk structure is presented in Fig. 5. In order to obtain this design we followed the same guidelines as in our previous work [8], but with some relaxed constraints thanks to the reduced length of our most recent optical waveguides. Details about the simulation techniques can hence be found in [8], where it is also shown that reduction of the motional amplitude at the clamping parts is the dominant mechanism for the Q increase. Fig. 5a shows the vibration profile of the upper disk's RBM. Fig. 5b presents clamping-loss limited mechanical $Q$ as a function of two key parameters: the pedestal radius and the thickness of the lower disk. The upper and lower disk radius are fixed to 1.05 µm, while the upper disk thickness is kept at 320 nm. The upper pedestal height is 500 nm and lower is 300 nm to minimize optical losses. The obtained new geometry appears promising: $Q$ reaches over 100 000 in a large parameter range (302 < t < 310 nm and 100 < r < 250 nm). These tolerances are met by our actual technological capabilities.

**Conclusion**

In conclusion, the use of a mechanical shield integrated in the pedestal is proved to enhance the $Q$ of GaAs disk optomechanical resonators by a factor more than 10. Further enhancement and easier fabrication are expected with a double-disk geometry, with $Q$ in excess of $10^5$ at GHz frequencies for technologically tractable geometries. Compared to in-plane phononic shield structures employed in optomechanical crystals [11], the shielded pedestal approach reported here presents a relatively simple design, together with a small footprint that opens the perspective of dense arrays of resonators. With their high circular symmetry, shielded disk resonators may push further our ability to strongly confine photons and phonons in an ultra-low dissipation optomechanical unit.

## Acknowledgments

This work was supported by the French ANR through the "Nomade" project, the European Research Council ERC trough the "Ganoms" project and the City of Paris through the "Research in Paris" program.